\documentclass[aps,pre,twocolumn]{revtex4}
\usepackage{bm}
\usepackage{graphicx}

\begin{document}

\title{Turbulent channel without boundaries: The periodic Kolmogorov flow}

\author{S. Musacchio}
\affiliation{Universit\'e de Nice Sophia Antipolis, CNRS, 
Laboratoire J.A. Dieudonn\'e, UMR 7351, 06100 Nice, France} 
\author{G. Boffetta}
\affiliation{Dipartimento di Fisica and INFN, Universit\`a di Torino, 
via P. Giuria 1, 10125 Torino, Italy}
\date{\today}

\begin{abstract}
The Kolmogorov flow provides an ideal instance of a virtual channel flow: 
It has no boundaries, but nevertheless it possesses 
well defined mean flow in each half-wavelength. 
We exploit this remarkable feature 
for the purpose of investigating the interplay 
between the mean flow and the turbulent drag of the bulk flow. 
By means of a set of direct numerical simulations 
at increasing Reynolds number
we show the dependence of the bulk turbulent drag 
on the amplitude of the mean flow. 
Further, we present a detailed analysis of the scale-by-scale energy balance, 
which describes how kinetic energy is redistributed among different regions of the flow 
while being transported toward small dissipative scales.   
Our results allow us to obtain an accurate prediction 
for the spatial energy transport at large scales. 
\end{abstract}
\maketitle 

\section{Introduction}
\label{sec:1}

Theoretical and numerical studies of turbulent flows can be 
divided into two categories. The first class of studies, mainly
motivated by experiments and practical applications, considers 
turbulence as generated by the interaction of the flow with a solid object. 
The simplest, and largely studied, example is the interaction 
with a plane, as in the turbulent channel.
The other category focuses mainly on intrinsic properties
of turbulence: bulk quantities which may be expected to give 
universal statistics independently on the way the 
flow is generated. These studies are usually based 
on the simplest possible geometry in the absence of boundaries, 
the so called homogeneous-isotropic turbulence in periodic domains. 

Between these two widely studied classes, there is another class of
inhomogeneous flow in the absence of boundaries. In these flows,
of which the Kolmogorov flow is the most studied example, 
homogeneity and isotropy are broken not by physical boundaries but
by the body force which generates the flow.
This flow was proposed by Kolmogorov as a model to understand 
the transition to turbulence, and was first studied by his students
who showed that the laminar solution becomes unstable 
to large scale perturbations at the critical Reynolds number $Re=\sqrt{2}$. 
Further studies investigated analytically the evolution
of the perturbation just above the instability \cite{S1985}
and numerically the transition to turbulence \cite{BO1996,She1987,Fortova2013}.
Because it is very convenient for analytical studies and 
numerical simulations, the 
Kolmogorov flow has been also used for several investigations
in anisotropic and/or inhomogeneous conditions, e.g. to investigate 
the anisotropy decompositions of turbulent flows \cite{SW1997,BT2010},
the instabilities in presence of Rossby waves \cite{LVF1999} 
stratification \cite{BY2002} and viscoelastic solution 
\cite{BCMPV2005,BBCMPV2007}.
Another important example is the Taylor-Green vortex which is closely
related to the von Karman flow used in experimental studies of 
hydrodynamics and MHD turbulence.
The different symmetries of these flows makes them suitable 
to investigate different classes of questions. 
The Taylor-Green flow is characterized by a shear 
region between two counter-rotating vortices 
and has been widely used in numerical studies of MHD dynamo 
(see e.g. \cite{ktvfb11,bbkmpr13})

The Kolmogorov flow can be thought as 
a simplified channel flow without boundaries. 
It displays a mean velocity profile which vanishes 
at the nodes of the sinusoidal force. Therefore it can be seen 
as a series of virtual channels, whose width is equal to half-period of
the forcing, flowing in alternate directions without being confined 
by material boundaries. 
On the other hand, because of the lack of boundaries, 
in the Kolmogorov flow the complex flow structures produced
by the wall, injected into the bulk and responsible for the energy 
transfer in bounded channel flows are missing. 
For these reasons the periodic Kolmogorov
allows to isolate bulk properties, e.g. of the turbulent drag,
which in a real channel flow might be hidden by the complex near-wall
phenomenology.
In this spirit it has been recently used to study the drag-reduction 
phenomenon induced by polymer additives \cite{BCM2005}.

The drag coefficient, or friction factor, is defined as the
ratio between the work made by the force and the kinetic energy carried 
by the mean flow. 
This fundamental, dimensionless number measures the power that 
has to be supplied to the fluid to maintain a given throughput. 
In general, when the flow is laminar, the drag coefficient is inversely 
proportional to the Reynolds number. Upon increasing
the intensity of the applied force the flow eventually becomes 
turbulent, and the drag coefficient becomes approximately independent 
of the Reynolds number $Re$ and therefore substantially larger than in 
the laminar case.

No exact values for the friction factor are know, even in 
simple geometry. In the case of smooth pipe flows, an empirical
logarithmic formula (Colebrook–White equation) reproduces 
accurately the experimental data. 
From a different perspective, 
rigorous mathematical bounds have been derived for the friction factor
with different geometries, also for the Kolmogorov flow \cite{CKG2001}.
In spite of their importance from a theoretical point of view, 
they are not strongly constrictive and therefore not very useful 
for applications.

Here we present the results of numerical simulations of the turbulent 
Kolmogorov flow aimed to study the dependence of turbulent drag 
on the Reynolds number. We also present a detailed analysis 
of the scale-by-scale energy balance 
which shows how the kinetic energy is redistributed among different regions 
and different scales of each virtual channel. Moreover, we will 
discuss the statistics of small scale velocity fluctuations 
and the scaling of structure functions in the inertial range of scales.

\section{Phenomenology of the Kolmogorov flow}
\label{sec:2}
We consider the Navier-Stokes equations for an incompressible 
velocity field $u_i({\bf x},t)$ ($i=1,2,3$) 
\begin{equation}
\partial_t u_i + u_j \partial_j u_i = - \partial_i p + \nu \partial^2 u_i 
+ g_i
\label{eq:2.1}
\end{equation}
forced by the Kolmogorov body force 
$g_i=\delta_{i,1} F \cos(z/L)$. Equation (\ref{eq:2.1}) admits a
stationary solution, the laminar Kolmogorov flow 
$u_i=\delta_{i,1} U_0 \cos(z/L)$ with $F=\nu U_0/L^2$.
This laminar solution becomes unstable to transverse large scale perturbations
(on scales much larger than $L$) when the Reynolds number 
$Re \equiv UL/\nu$ exceed the threshold $Re_c=\sqrt{2}$ \cite{MS1961}.
While this instability is two-dimensional (Squire theorem), by
increasing $Re$ the flow develops further instabilities and 
eventually becomes three-dimensional and turbulent.
Here we consider the case $Re \gg Re_c$ for which linear and
weakly non-linear analysis is not applicable and therefore we
will make use of Direct Numerical Simulations (DNS) of (\ref{eq:2.1}).

An interesting property of the Kolmogorov flow is that even in
the turbulent regime, the mean velocity
has nearly the Kolmogorov profile \cite{BO1996}
\begin{equation}
\overline{u_1({\bf x},t)} = U \cos(z/L)
\label{eq:2.2}
\end{equation}
where the overbar denotes space-time average over $x$, $y$ and $t$.
Moreover, the Reynolds stress is also monochromatic
\begin{equation}
\overline{u_1 u_3} = S \sin(z/L)
\label{eq:2.3}
\end{equation}
with amplitude $S$ 
and therefore the momentum budget (obtained by averaging (\ref{eq:2.1}))
becomes a simple algebraic relation for the coefficients
of the monochromatic terms
\begin{equation}
F = {S \over L} + {\nu U \over L^2}
\label{eq:2.4}
\end{equation}

The friction coefficient $f$ for the Kolmogorov flow can be defined as
ratio between the work done by the force and the kinetic energy 
of the flow
\begin{equation}
f=FL/U^2
\label{eq:2.5}
\end{equation}
which, because the energy input is simply 
$\epsilon = \langle u_i f_i \rangle = {1 \over 2} F U$
($\langle ... \rangle$ represents average over the whole space),
is equivalent also to the dissipation factor
\begin{equation}
f = {2 \epsilon L \over U^3}
\label{eq:2.6}
\end{equation}
We observe that in literature the dissipation factor 
is sometimes defined in terms of the root mean square (rms) velocity 
$U_{rms}=\langle |{\bf u}|^2 \rangle^{1/2}$ as $\beta=\epsilon L/U_{rms}$
\cite{DES2003}.
Numerical simulations shows that $U_{rms}$ is proportional to $U$ (see
below) and so are therefore $f$ and $\beta$, but an explicit relation 
between the two dimensionless coefficient is not known. 
Together with the friction factor, we define also the dimensionless stress
coefficient $\sigma \equiv {S/U^2}$
and therefore we can rewrite the momentum budget (\ref{eq:2.4}) as
\begin{equation}
f = \sigma + {1 \over Re}
\label{eq:2.7}
\end{equation}

In the laminar fix point, for which $S=0$ we have from (\ref{eq:2.7})
the usual laminar expression for the friction factor
\begin{equation}
f_{lam} = {1 \over Re}
\label{eq:2.8}
\end{equation}
As $Re$ increases, the laminar solution becomes unstable and 
the friction factor becomes larger than $f_{lam}$ and eventually
approaches a constant as $Re >> 1$. This corresponds to the 
so-called 
"zeroth law of turbulence" \cite{Frisch1995}.

From a mathematical point of view, although $f$ cannot be computed 
analytically in a turbulent flow, several bounds have been obtained.
The simplest lower bound is given by the laminar expression 
(\ref{eq:2.7}), which corresponds to the absence of turbulence.
In the case of Kolmogorov flow with periodic 
boundary conditions an upper bound for the dissipation factor 
in the limit of high Reynolds numbers is \cite{RDD2011}
$\beta \le \beta_b = \pi/\sqrt{216} \simeq 0.214$.

\section{Results from numerical simulations}
\label{sec:3}

We integrated (\ref{eq:2.1}) on a cubic periodic box of size
$L_{box}=2 \pi$ with Kolmogorov forcing at scale $L=1$,
fixed viscosity and different values of forcing amplitude $F$. 
Starting from zero velocity configuration, a turbulent, statistically 
stationary state, is reached after several large-scale eddy turnover
times $T$. 
The value of $F$ determines the amplitude of the velocity in the 
flow and therefore the Reynolds number as shown in Table~\ref{table1}.
As $Re \gg Re_{c}$ the flow is always in the turbulent regime.

After the flow has reached a stationary condition, we compute the mean 
profiles from which we obtain $U$ and $S$ by fitting with (\ref{eq:2.2})
and (\ref{eq:2.3}) and we also measure the other statistical properties
of the flow.
We remark that the use of a forcing at the smaller wavenumber 
generates strong fluctuations in the large scale properties of the flow, 
therefore we have to average over many $T$ (between $10$ and $100$) to 
have a good convergence of mean quantities.
We check the convergence to a statistical steady state by using 
(\ref{eq:2.7}) which is indeed satisfied with good accuracy. 

\begin{table}[t!]
\begin{tabular}{c|c|c|c|c|c|c|c}
$Re$ & $F$ & $U$ & $u'_{rms}$ & $\epsilon$ & $\eta$ & $\tau_{\eta}$ & $T$ \\ \hline
$60$ & $0.0005$ & $0.060$ & $0.032$ & $1.49 \times 10^{-5}$ & $9.06 \times 10^{-2}$ & $8.21$ & $129$ \\
$78$ & $0.001$ & $0.078$ & $0.042$ & $3.93 \times 10^{-5}$ & $7.10 \times 10^{-2}$ & $5.05$ & $93.1$ \\
$120$ & $0.002$ & $0.12$ & $0.062$ & $1.17 \times 10^{-4}$ & $5.40 \times 10^{-2}$ & $2.92$ & $68.2$ \\
$160$ & $0.004$ & $0.16$ & $0.087$ & $3.16 \times 10^{-4}$ & $4.22 \times 10^{-2}$ & $1.78$ & $50.2$ \\
$230$ & $0.008$ & $0.23$ & $0.12$ & $9.31 \times 10^{-4}$ & $3.22 \times 10^{-2}$ & $1.04$ & $36.3$ \\
$340$ & $0.016$ & $0.34$ & $0.17$ & $2.70 \times 10^{-3}$ & $2.47 \times 10^{-2}$ & $0.61$ & $25.2$ \\
$480$ & $0.032$ & $0.48$ & $0.25$ & $7.73 \times 10^{-3}$ & $1.90 \times 10^{-2}$ & $0.36$ & $18.4$ \\
$730$ & $0.064$ & $0.73$ & $0.38$ & $2.30 \times 10^{-2}$ & $1.44 \times 10^{-2}$ & $0.21$ & $13.1$ \\
$990$ & $0.128$ & $0.99$ & $0.53$ & $6.41 \times 10^{-2}$ & $1.12 \times 10^{-2}$ & $0.13$ & $9.36$ \\
$1350$ & $0.256$ & $1.35$ & $0.76$ & $1.73 \times 10^{-1}$ & $8.72 \times 10^{-3}$ & $0.076$ & $6.72$ \\
$2000$ & $0.512$ & $2.00$ & $1.08$ & $5.23 \times 10^{-1}$ & $6.61 \times 10^{-3}$ & $0.044$ & $4.78$ 
\end{tabular}
\caption{Parameters of the simulations. $F$ amplitude of the 
forcing, $U$ amplitude of the mean
profile, $Re=U L/\nu$, $u'_{rms}$ rms of the fluctuation of the $x$
component of the 
velocity, $\epsilon=\nu \langle ({\bf \partial} {\bf u})^2 \rangle$
mean energy dissipation, $\eta=(\nu^3/\epsilon)^{1/4}$ Kolmogorov scale,
$\tau_{\eta}=(\nu/\epsilon)^{1/2}$ Kolmogorov timescale, 
$T=\langle {\bf u}^2 \rangle/(2 \epsilon)$ large scale time.
The integral scale $L=1$ and the viscosity $\nu=10^{-3}$ are 
fixed for all simulation. Simulations up to $Re=480$ are done at
resolution $N=128$, $Re=730$ and $Re=990$ with $N=256$ and 
$Re=1350$ and above with $N=512$. For all the simulations
$k_{max} \eta \ge 1$.}
\label{table1}
\end{table}
\begin{figure}[ht!]
\includegraphics[width=9cm]{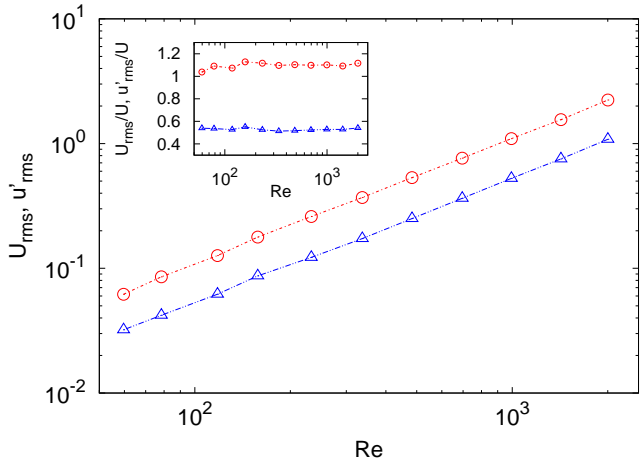}
\caption{Rms velocity $U_{rms}$ (red circles) 
and turbulent fluctuations velocity $u'_{rms}$ (blue triangles)
as a function of the Reynolds number $Re=UL/\nu$. 
The inset shows the two rms velocities normalized 
by the mean velocity amplitude $U$.}
\label{fig1}
\end{figure}

A first remarkable result obtained from our simulations concerns the 
intensities of the turbulent fluctuations at different Reynolds numbers.
We decompose the flow
in the mean velocity and fluctuations as 
$u_i({\bf x},t)=\bar{u_i}(z)+u'_i({\bf x},t)$ (where $\bar{u_1}$ is given
by (\ref{eq:2.2}) and $\bar{u_2}=\bar{u_3}=0$). 
Figure~\ref{fig1} shows that the rms turbulent fluctuations $u'_{rms}$ 
grows linearly with $Re$ and it is proportional to the 
mean velocity amplitude $U$. In particular we 
obtain $u'_{rms}/U \simeq 0.54 \pm 0.03$
in the range of $Re$ investigated. 
The same behavior is observed for the rms velocity $U_{rms}$.  
We find  $U_{rms}/U \simeq 1.10 \pm 0.02$. 
This result confirms that the friction factors 
$f$ and $\beta$, which are defined on the basis of $U$ and 
$U_{rms}$ respectively, are proportional to each other, 
as we anticipated in the previous section. 
Of course we expect different ratios $u'_{rms}/U$ and $U_{rms}/U$ 
for much smaller values of $Re$, 
close to the instability threshold.

\subsection{Momentum budget}
\label{sec:3.1}

\begin{figure}[ht!]
\includegraphics[width=9cm]{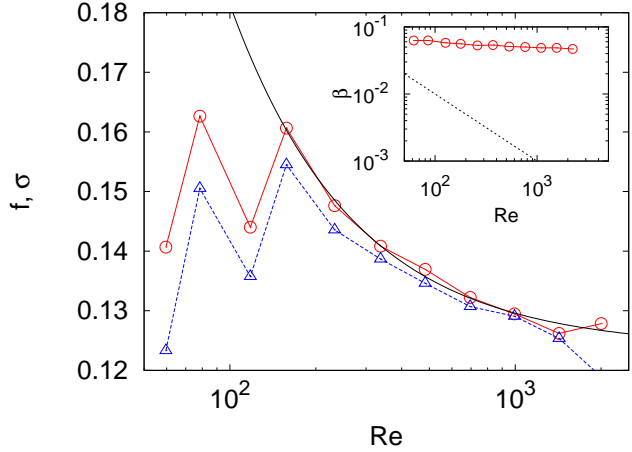}
\caption{Evolution of the friction coefficient $f=FL/U^2$ (red circles) 
and stress coefficient $\sigma=S/U^2$ (blue triangles) 
as a function of the Reynolds number $Re=UL/\nu$. 
The black line represents the fit with (\ref{eq:3.1}) which 
gives $f_0=0.124$ and $b=5.75$.
Inset: Dissipation factor $\beta=\epsilon L/U_{rms}$ 
vs $Re=U_{rms}L/\nu$ for the set of simulations. 
The dashed line represents the laminar lower bound
$\beta_{lam}=1/Re$.}
\label{fig2}
\end{figure}

In Fig.~\ref{fig2} we show the friction 
coefficient $f =FL/U^2$ and the stress coefficient $\sigma=S/U^2$
as a function of $Re$ as obtained from the numerical simulations.
We find that, for $Re \gtrsim 160$, the friction coefficient follows  
with good approximation 
\begin{equation}
f = f_0 + {b \over Re}
\label{eq:3.1}
\end{equation}
and therefore from (\ref{eq:2.7}) 
\begin{equation}
\sigma = f_0 + {b-1 \over Re}
\label{eq:3.2}
\end{equation}
The fit for $f$ with (\ref{eq:3.1}) gives $f_0=0.124$.
It is interesting to note that 
in the Kolmogorov flow the asymptotic behavior 
(\ref{eq:3.1},\ref{eq:3.2}),
which describes the large-$Re$ limit, 
is already present for relatively small $Re$. 
In our set of simulations it can be observed for $Re \gtrsim 160$, 
which corresponds to the onset of the fully developed turbulence regime,  
as we will show in Section~\ref{sec:3.4}. 
This is at variance with the case of pipe flows, 
in which the asymptotic behavior of the drag coefficient
appears at much larger $Re$, 
and the laminar regime is still present  
for $Re$ in the range investigated in our study. 
We remind that in pipe flows the laminar regime is linearly stable, 
while the Kolmogorov flow becomes linearly unstable already 
at $Re>\sqrt{2}$. 
The early manifestation of the large-$Re$ asymptotic regime 
in the Kolmogorov flow is important
from the point of view of application, because it justifies the 
extrapolation of large-$Re$ behavior from relatively low-$Re$ simulations.

The dissipation factor $\beta$ is shown in the inset of Fig.~\ref{fig2} 
as a function 
of the Reynolds number, here defined in terms of $U_{rms}$ for consistency
with previous literature. 
While a weak dependence on $Re$ is still observable, numerical data suggests
an asymptotic value, as $Re \to \infty$, $\beta \lesssim 0.05$, 
consistent but quite smaller than the bound $\beta_b \simeq 0.214$.

\subsection{Local energy balance}
\label{sec:3.2}

In stationary condition we can write, by multiplying (\ref{eq:2.1})
by $u_i$ and by averaging over $(x,y)$, the energy balance profile
\begin{equation}
\epsilon_I(z) \equiv \overline{u_i f_i} = \epsilon_{\nu}(z) + T(z)
\label{eq:3.2.1}
\end{equation}
where the energy dissipation profile is 
\begin{equation}
\epsilon_{\nu}(z) \equiv \nu \overline{|{\bf \nabla} {\bf u}|^2}
\label{eq:3.2.2}
\end{equation}
and $T(z)$ is
\begin{equation}
T(z) = 
\partial_z \overline{u_3 \left(u^2/2 + p \right)}
- \nu \partial_z^2 \overline{u^2/2}
\label{eq:3.2.2b}
\end{equation}
Given the monochromatic mean profile for the velocity field, we have
\begin{equation}
\epsilon_{I}(z) = {F U \over 2} \left[1+ \cos(2 z/L) \right]
\label{eq:3.2.3}
\end{equation}
with average $\epsilon_I=FU/2=F^{3/2}(L/4 f)^{1/2}$.

In (\ref{eq:3.2.1}) $T(z)=\partial_z J(z)$
represents the spatial energy transport which can be $T(z)>0$ where the 
energy is locally injected and $T(z)<0$ where it is removed. Of course
$\langle T(z) \rangle_z=0$ for energy conservation (and 
$\epsilon_I=\epsilon_{\nu}$).

\begin{figure}[ht!]
\includegraphics[width=9cm]{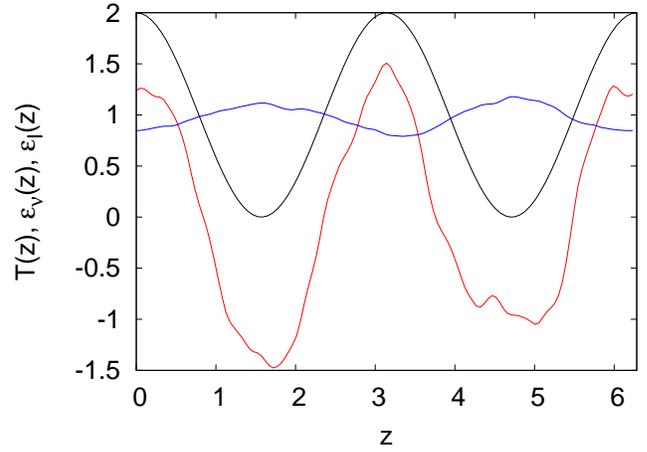}
\caption{Profiles of energy dissipation $\epsilon_{\nu}(z)$ (blue line),
energy transport $T(z)$ (red line) which sum to energy input
$\epsilon_I(z)$ (black line) according to (\ref{eq:3.2.1}). 
All quantities have been normalized to the mean energy input 
$\epsilon_I=FU/2$. Data from simulation at $Re=160$.}
\label{fig3}
\end{figure}

\begin{figure}[ht!]
\includegraphics[width=9cm]{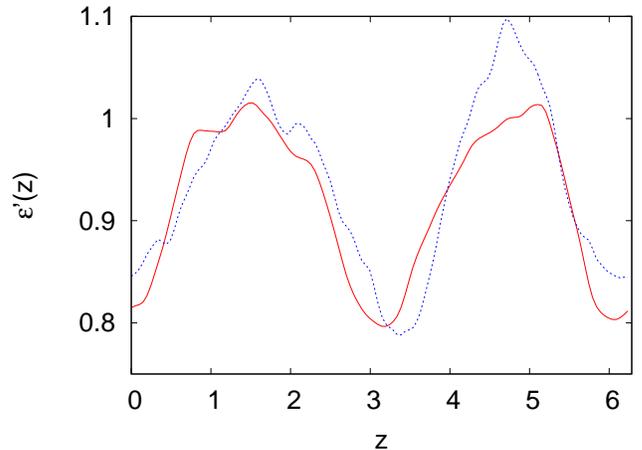}
\caption{Profiles of local dissipation of fluctuation energy
$\epsilon'(z)=\nu \overline{(\partial_j {u'}_i)^2}$ normalized
with the mean energy input $FU/2$ for $Re=78$ (red continuous line) and 
$Re=160$ (blue dotted line).}
\label{fig4}
\end{figure}

Figure~\ref{fig3} shows the different terms in (\ref{eq:3.2.1}). 
Because the dissipation term is almost homogeneous, the transfer term
mainly reflects the profile of the energy input (\ref{eq:3.2.3}). 
A small modulation is observable in the dissipation which is found to
be almost independent on the Reynolds number \cite{BO1996}. 
By means of a Reynolds decomposition of the velocity field in the
mean profile and fluctuating components: 
$u_i({\bf x},t)=\overline{u_i({\bf x},t)}+u'_i({\bf x},t)$
(where $\overline{u_i}\ne 0$ for $i=1$ only)
the energy dissipation (\ref{eq:3.2.2}) can be rewritten as
\begin{equation}
\epsilon_{\nu}(z) = \nu {U^2 \over L^2} \sin^2(z/L) +
\nu \overline{(\partial_j u'_i)^2}
\label{eq:3.2.4}
\end{equation}
where we have used (\ref{eq:2.2}).
The first term in (\ref{eq:3.2.4}) represents the direct dissipation
by viscosity on the large scale mean flow which, when normalized with
the mean energy input $F U/2$, decays as $1/Re$. 
The second term in (\ref{eq:3.2.4}) represents the local dissipation
of velocity fluctuations. Its contribution in the energy balance,
proportional to velocity gradient, is weakly dependent on $Re$ and it
is still inhomogeneous in $z$, as shown in Fig.~\ref{fig4}.
The term is responsible for the modulation observed in Fig.~\ref{fig3}
which persists also for larger Reynolds numbers.

If we neglect these small modulations, and assuming as a zero-order 
approximation an homogeneous dissipation
$\epsilon_{\nu}(z)=\epsilon_{\nu}=FU/2$, from (\ref{eq:3.2.1}) we
obtain an explicit expression for the energy transport term
\begin{equation}
T(z) = {F U \over 2} \cos(2 z/L)
\label{eq:3.2.5}
\end{equation}

\subsection{Spatial and scale dependence of the energy flux}
\label{sec:3.3}

We have seen in the previous Section that the energy dissipation profile
$\epsilon_{\nu}(z)$ (at small scales) is much more homogeneous than
the energy input profile $\epsilon_I(z)$ (at large scales).
This means that the energy flux which, on the average, transfers
energy from large to small scales, also redistribute energy in 
space. It is therefore interesting to investigate how the energy
is transfer at different position $z$ at the different scales in the
turbulent cascade.

In order to get more insight in this mechanism of energy transfer, we
consider the scale by scale budget of kinetic energy 
\cite{Germano1992,Eyink1995}. We introduce a filter kernel 
$G_{\ell}({\bf x})=\ell^{-3} G({\bf x}/\ell)$ 
(with $\int d^3 x G({\bf x})=1$) which defines a low-passed
filtered field by the convolution 
$u^{(\ell)}_{i}({\bf x})\equiv (G_{\ell} \star u_i)({\bf x})$. 
By applying the filter to the equation of motion (\ref{eq:2.1}),
contracted with $u_i$, we get
the equation for the energy at large scale
\begin{equation}
\partial_t E^{(\ell)}({\bf x}) + \partial_j J^{(\ell)}_{j}({\bf x})=
-\Pi^{(\ell)}({\bf x})- D^{(\ell)}({\bf x}) + F^{(\ell)}({\bf x})
\label{eq:3.3.1}
\end{equation}
where $E^{(\ell)}=(1/2) |{\bf u}^{(\ell)}|^2$ is the large scale kinetic energy
density, $D^{(\ell)}=\nu |{\bf \nabla} {\bf u^{(\ell)}}|^2$ is large scale
energy dissipation, 
$J^{(\ell)}_{i}=u^{(\ell)}_{j}[\tau^{(\ell)}_{ij}+\delta_{ij} 
(E^{(\ell)}+p^{(\ell)})]-\nu \partial_i E^{(\ell)}$
the spatial energy transport in the large scales and 
\begin{equation}
\Pi^{(\ell)}({\bf x})=-\tau^{(\ell)}_{ij} \partial_j u^{(\ell)}_{i}
\label{eq:3.3.2}
\end{equation}
is the scale to scale energy flux, 
where $\tau^{(\ell)}_{ij}=(u_i u_j)^{(\ell)} - u^{(\ell)}_{i} u^{(\ell)}_{j}$
is the stress tensor (of the filtered field).
The term $\Pi_{\ell}({\bf x})$ represents the local energy
flux to scales smaller than $\ell$ at point ${\bf x}$. 

\begin{figure}[ht!]
\includegraphics[width=9cm]{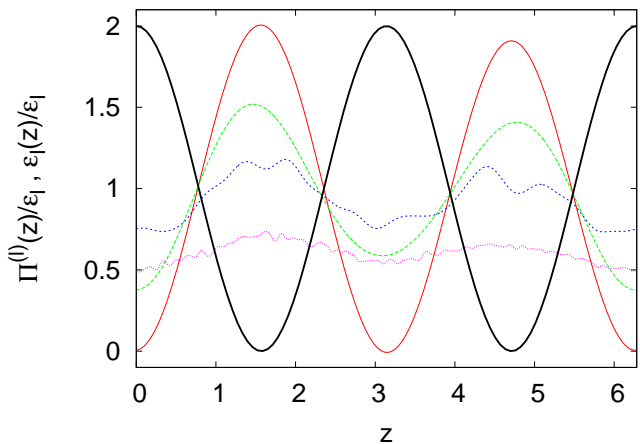}
\caption{Profile of the scale-to-scale energy flux $\overline{\Pi^{(\ell)}}(z)$
for different scales of the filter $\ell=L/2$ (red continuous line),
$\ell=L/4$ (green dashed line), $\ell=L/16$ (blue dotted line) and
$\ell=L/64$ (pink dotted-dashed) for the run at $Re=2000$.
The black line represents the theoretical energy input
$\epsilon_I(z)=FU \cos^2(z/L)$.
All quantities are normalized with the mean energy input $FU/2$.}
\label{fig5}
\end{figure}
In our setup we are interested to the horizontally averaged version
of (\ref{eq:3.3.1}) which, in stationary conditions and for $\ell$ smaller 
than the forcing scale ($\ell < L$), reads
\begin{equation}
\partial_z \overline{J^{(\ell)}_{3}}(z) = - \overline{\Pi^{(\ell)}}(z) -
\overline{D^{(\ell)}}(z)+\epsilon_I(z)
\label{eq:3.3.3}
\end{equation}
When averaged over $z$, the first term in (\ref{eq:3.3.3}) vanishes and
one obtain the homogeneous balance for the energy flux
\begin{equation}
\langle \Pi^{(\ell)} \rangle = - \langle D^{(\ell)} \rangle + \epsilon_I 
\label{eq:3.3.4}
\end{equation}
In the inertial range of scales ($L \gg \ell \gg \eta$) the dissipative 
term is negligible and one has $\langle \Pi^{(\ell)} \rangle =\epsilon_I$.
This mean flux is reduced at smaller scales by the presence of dissipation.
Figure~\ref{fig5} shows the energy transport profile 
$\overline{\Pi^{(\ell)}}(z)$ for different values of the filter 
scale $\ell$, together with the energy input $\epsilon_I(z)$.
At the largest scale ($\ell=L/2$) the flux is strongly inhomogeneous,
while moving to smaller scales it becomes more uniform. In the inertial
range of scales the $z$-averaged value of $\overline{\Pi^{(\ell)}}(z)$
is constant (and equal to the input), as shown by the first three 
curves in Fig.~\ref{fig5}. Moving to smaller scales, closer to the 
dissipative range of scales, the term $D^{(\ell)}(z)$ in (\ref{eq:3.3.3})
is not negligible any more and consequently the average flux decreases.

We find that the profile of the scale-to-scale flux is never 
negative and vanishes in correspondence of the maximum input
at the largest scale. This means that there is no back-scatter of
energy at a given $z$ in the Kolmogorov channel  (while, of course, 
the one-point flux (\ref{eq:3.3.2}) can be negative).
This remarkable result
(which is found to be independent on the Reynolds number)
suggest that at the largest scale the z-averaged energy transport can
be simply expressed as 
$\overline{\Pi^{(\ell)}}(z) \simeq 2 \epsilon_I-\epsilon_I(z)$.
Using (\ref{eq:3.2.3}) and (\ref{eq:3.3.3}) one ends with a simple
prediction for the profile of the spatial transport at the largest scale 
\begin{equation}
\overline{J^{(L)}_{3}}(z) = {F U L \over 2} \sin(2z/L)
\label{eq:3.3.5}
\end{equation}

This result has a clear interpretation: $\overline{J^{(\ell)}_{3}}(z)$
represents the {\it current} of energy in the $z$ direction.
As it redistributes energy among different regions in the channel,
it is positive (to larger $z$) in regions where the input 
decreases ($\partial_z \epsilon_I(z)<0$) and negative (to smaller
$z$) in regions where the input increases ($\partial_z \epsilon_I(z)>0$).

\subsection{Structure functions}
\label{sec:3.4}

In the inertial range of scales, $\eta \ll \ell \ll L$, 
the amplitude of turbulent velocity fluctuations 
is expected to exhibit a scaling behavior.
The scaling behavior of the structure functions 
$S_p(\ell) = \langle (\delta u_{\ell})^p \rangle \sim \ell^{\zeta_p}$
encodes relevant informations on the statistics of  
longitudinal velocity increments 
$\delta u_{\ell} = [\bm{u}(\bm{x}+\bm{\ell})-\bm{u}(\bm{x})]\cdot\hat{\bm{\ell}}$.
 
In the case of the Kolmogorov flow
the structure functions are expected to show a dependence 
on the coordinate $z$ and the direction $\hat{\bm{\ell}}$, 
due to the inhomogeneity and anisotropy of the forcing. 
In order to extract their homogeneous and isotropic projection
we have averaged the structure functions 
over several isotropically distributed directions $\hat{\bm{\ell}}$
and over all the available values of the $z$ coordinate. 

\begin{figure}[ht!]
\includegraphics[width=9cm]{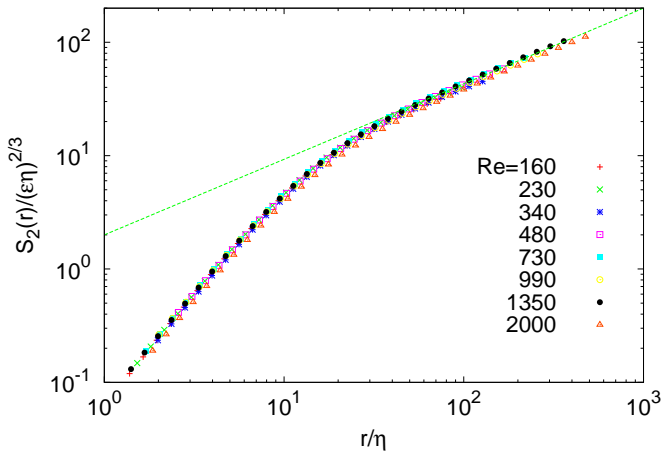}
\caption{Isotropic second-order structure function $S_2(\ell)$ for 
various $Re$.  The dashed line is the dimensional scaling 
$S_2(\ell) \sim (\varepsilon \ell)^{2/3}$.}
\label{fig6}
\end{figure}
In Figure~\ref{fig6} we show the second-order structure function 
$S_2(\ell)$ obtained from our simulations at various $Re$.
We observe a good agreement with the dimensional scaling 
$S_2(\ell) \sim (\varepsilon \ell)^{2/3}$ and 
a remarkable collapse of the curves 
when the scales $\ell$ are normalized with the Kolmogorov scale $\eta$
and the amplitude of $S_2(\ell)$ is rescaled with the dimensional factor
$(\varepsilon \eta)^{2/3}$. 
The limited scaling range does not allow us to investigate the presence 
of intermittency corrections. 

\begin{figure}[ht!]
\includegraphics[width=9cm]{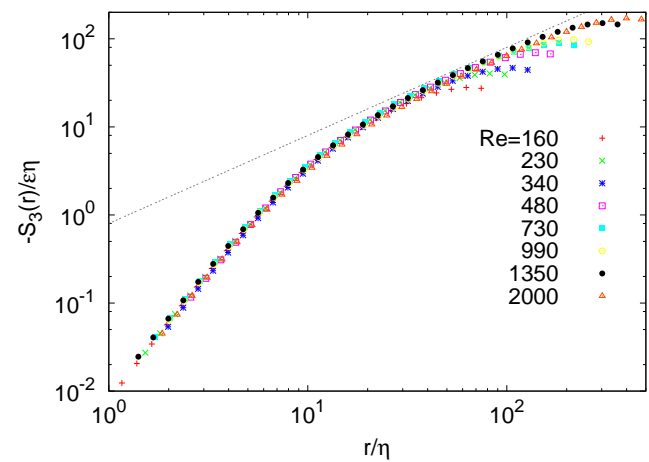}
\caption{Isotropic third-order structure function $S_3(\ell)$ 
for various $Re$. 
The dashed line represents the $4/5th$ law 
$S_3(\ell) = -4/5 \varepsilon \ell $.}
\label{fig7}
\end{figure}

\begin{figure}[ht!]
\includegraphics[width=9cm]{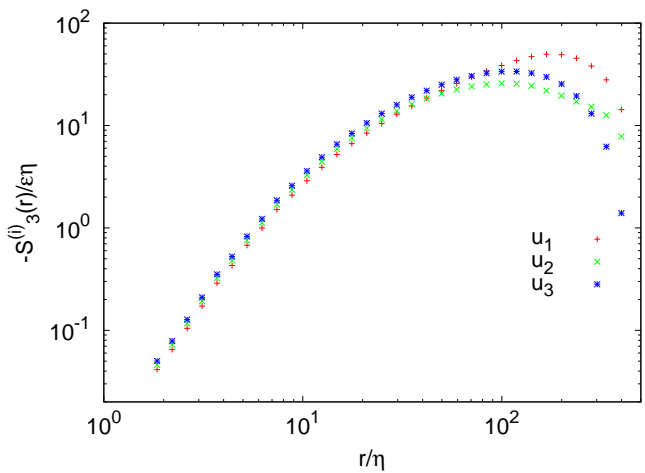}
\caption{Longitudinal third-order structure function $S^{(i)}_3(\ell)$ 
in the direction $i=1,2,3$ for the run at $Re=2000$.}
\label{fig8}
\end{figure}
The negative sign of the third-order structure functions $S_3(\ell)$
in the inertial range signals 
the direction of the mean energy transfer from large to small scales. 
The scaling behavior (shown in Figure~\ref{fig7}) is in good agreement 
with the Kolmogorov $4/5th$ law
$S_3(\ell) = -(4/5) \varepsilon \ell$. 
It is interesting to note that 
$S_3(\ell)$ begins to exhibit a scaling range at $Re\simeq 160$, 
which is the lowest $Re$ at which the friction factor 
$f$ begins to display the asymptotic behavior $f \simeq f_0 + b/Re$
(see Figure~\ref{fig2}). 
Indeed the ``zeroth law of turbulence'', i.e. the fact that 
the friction coefficient becomes almost constant as $Re \to \infty$, 
is strictly connected to the development of the turbulent energy cascade. 

In Figure~\ref{fig8} we show the longitudinal
third-order structure functions $S^{(i)}_3(\ell)$
computed along the directions of the axes $i=1,2,3$.
The behavior in the different directions reflects the anisotropy of the flow.  
In particular the $S_3$ in the forced ($i=1$) direction displays a 
broader scaling range with respect to the other two, which 
is consistent with the fact that the flow in this direction is more
energetic.

\section{Conclusions}
\label{sec:4}

In this paper we have presented an 
analysis of the momentum end energy balance 
of the turbulent Kolmogorov flow. 
By means of a set of direct numerical simulations 
at increasing Reynolds number we have shown that rms value of turbulent 
fluctuations $u'_{rms}$ grows linearly with the amplitude of the mean flow $U$, 
and that the friction coefficient $f=FL/U^2$ follows the asymptotic behavior 
$f= f_0 + O(Re^{-1})$ as $Re \to \infty$.  

We have shown that the local flux of kinetic energy 
has a strong dependence both on the scale $\ell$ 
on the vertical coordinate $z$. 
The maximum energy flux toward small dissipative scales occurs  
at the nodes of the Kolmogorov flow, i.e. the regions located 
on the vertical positions where the mean flow is vanishing and 
the mean shear is maximum. 
Conversely, the minimum energy flux is observed at 
the antinodes where the mean flow is maximum. 
The amplitude of this spatial modulation of the energy flux reduces 
as the turbulent cascades proceeds toward small scales, 
but it is still present at dissipative scales. 

We have also derived a prediction for the spatial transport of kinetic energy,
which describes how kinetic energy is redistributed among different regions of
the flow while being transported toward small dissipative scales.  In
particular we have shown that there is an energy current from the antinodes to
the nodes which transports kinetic energy from the regions where the energy
input provided by the forcing is maximum, i.e the maxima of the mean flow,
toward the regions where the input vanishes, i.e. the maxima of the mean shear.
As a consequence, this current produces a partial recovery of the homogeneity
of the flow. 

From a theoretical point of view, the Kolmogorov flow represent an ideal
framework to investigate the properties of spatial transfer of kinetic energy
in non-homogeneous, sheared turbulent flows.  In spite of the absence of
material boundaries, it allows to define mean profiles for all the relevant
quantities, e.g, the mean velocity, the mean shear, the turbulent fluxes.
Thanks to this remarkable feature it can be used to investigate the interplay
between the mean flow and the bulk turbulence,  avoiding at the same time the
complexities induced by the development of turbulent boundary layers.  It
provides therefore an ideal tool to study the properties of internal shears in
turbulent flows, which appears, e.g, in geophysical currents and jets.
In view of possible geophysical applications it would be very interesting
to investigate Lagrangian properties of the Kolmogorov flow, such as 
the absolute and relative dispersion of tracers.

Computer time provided by Cineca is gratefully acknowledged.

\bibliography{biblio}

\end{document}